\documentclass{ws-ijmpcs}
\usepackage{bbm}
\usepackage{graphicx}
\usepackage{epsfig}
\usepackage{bm}
\usepackage{amsfonts}
\usepackage{epstopdf}

\begin{document}

\markboth{Chao-Jun Feng}
{Instructions for Typing Manuscripts (Paper's Title)}

%
\catchline{}{}{}{}{}
%

\title{Quantum Spring}

\author{Chao-Jun Feng}

\address{Shanghai United Center for Astrophysics (SUCA), \\ Shanghai Normal University,
    100 Guilin Road, Shanghai 200234, P.R.China\\
fengcj@shnu.edu.cn}

\author{Xin-Zhou Li}

\address{Shanghai United Center for Astrophysics (SUCA), \\ Shanghai Normal University,
    100 Guilin Road, Shanghai 200234, P.R.China\\
kychz@shnu.edu.cn}

\maketitle

\begin{history}
\received{Day Month Year}
\revised{Day Month Year}
\end{history}

\begin{abstract}
In this paper, we will give a short review on \textit{quantum spring}, which is a Casimir effect from the helix boundary condition that proposed  in our earlier works. The Casimir force parallel to the axis of the helix behaves very much like the force on a spring that obeys the Hooke's law when the ratio $r$ of the pitch to the circumference of the helix is small, but in this case, the force comes from a quantum effect, so we would like to call it \textit{quantum spring}.  On the other hand, the force perpendicular to the axis decreases monotonously with the increasing of the ratio $r$. Both forces are attractive and their behaviors are the same in two and three dimensions.

\end{abstract}

\ccode{PACS numbers: 03.70.+k, 11.10.-z}

\section{Introduction}\label{sec:intro}
The Casimir effect firstly founded by Casimir \cite{Casimir:1948dh} has been extensively studied \cite{Plunien:1986ca} for more than 60 years. Essentially, the casimir effect is a polarization of the vacuum of some quantized fields, and it may be thought of as the energy due to the distortion of the vacuum. Such a distortion may be caused either by the presence of boundaries in the space-time manifold or by some background field like the gravity. Early works on the gravity effect were performed by Utiyama and DeWitt\cite{Utiyama:1962sn,DeWitt:1975ys}. In history, Casimir firstly predicts the effect of the boundaries and he found that there is an attractive force acting on two conducting plan-parallel plates in vacuum, and this phenomenon inspires much theoretical interest as macroscopic manifestation of quantum fluctuation. Since the last decade, the Casimir effect has been paid more attention due to the development of precise measurements \cite{Decca:2007yb}, and it has been applied to the fabrication of microelectromechanical systems (MEMS)\cite{MEMS}. Recently, some new methods have developed for computing the Casimir energy between a finite number of compact objects \cite{Emig:2007cf}.

The nature of the Casimir force may depend on (i) the background field, (ii) the spacetime dimensionality, (iii) the
type of boundary conditions, (iv) the topology of spacetime, (v) the finite temperature. The most evident example of
the dependence on the geometry is given by the Casimir effect inside a rectangular box \citeup{Plunien:1986ca,Lukosz}.
The detailed calculation of the Casimir force inside a D-dimensional rectangular cavity was shown in Ref.\refcite{Li},\refcite{Li2}, in
which the sign of the Casimir energy depends on the length of the sides. The Casimir force arises not only in the
presence of material boundaries, but also in spaces with nontrivial topology. For example, we get the scalar field on a
flat manifold with topology of a circle $S^1$. The topology of $S^1$ causes the periodicity condition
$\phi(t,0)=\phi(t,C)$, where $C$ is the circumference of $S^1$, imposed on the wave function which is of the same kind
as those due to boundary. Similarly, the antiperiodic conditions can be drawn on a M\"obius strip. The $\zeta$-function
regularization procedure is a very powerful and elegant technique for the Casimir effect. Rigorous extension of the
proof of Epstein $\zeta$-function regularization has been discussed in Ref.\refcite{Elizalde}. Vacuum polarization in the
background of on string was first considered in Ref.\refcite{Helliwell:1986hs}. The generalized $\zeta$-function has many
interesting applications, e.g., in the piecewise string \cite{Li:1990bz,Li:1990bz2}. Similar analysis has been applied to
monopoles \cite{BezerradeMello:1999ge}, p-branes \cite{Shi:1991qc} or pistons \cite{Zhai}\cdash\cite{Zhai4}.

It is well known that there are many things that look like the spring, for instance, DNA has the helix structure in our
cells. Thus, it is interesting to find the effect of the helix configuration presenting in the space-time manifold for quantum fields and we have found that the behavior of the force parallel to the axis of the helix is very much like the force on a spring that obeys the Hooke's law in mechanics when the $r\ll1$, which is the ratio of the pitch $h$ to the circumference $a$ of the helix. However, in this case, the force comes from a quantum effect, and so we would like to call the helix structure as a \textit{quantum spring} \cite{Feng:2010qj}\cdash\cite{Feng:2011}. We also generalized it to higher dimensions and also considered  the anti-boundary conditions imposed on a quantum field.

This paper is organized as follows. In next section we shall present a topological view of helix configuration and show the calculation results of the vacuum energy density and its corresponding Casimir force in Sec.\ref{sec:casimir}, and in the last section, we will give some conclusion and discussion.

\section{Topology of the flat (D+1)-dimensional spacetime}

The Casimir effect arise not only in the presence of material boundaries, but also in spaces with nontrivial topology.  The simplest example of the Casimir effect of topological origin is the scalar field on a flat manifold with topology $S^1$ as we mentioned before. Before we consider complicated cases in the flat spacetime, it is worth noting that the concept of quotient topology is very useful for concrete application and then we shall present a topological view of helix configuration. Now, we consider a surjective mapping $f$ from a topological space $X$ onto a set $Y$. The quotient topology on $Y$ with respect to $f$ is given in Ref.~\refcite{Mukres}. Surjective mapping can be easily obtained when we use the equivalence classes of some equivalence relation $\sim$. Thus, $X/\sim$ is defined to be the set of equivalence classes of elements of $X$, and the mapping $f: X\rightarrow X/\sim$ could be denoted by  $f(x) = [x]$, which is the equivalence class containing $x$. For example, consider the unit square $I^2 = [0,1]\times [0,1]$ and the equivalence relation $\sim$ generated by the requirement that all boundary points be equivalent, thus identifying all boundary points to a single equivalence class. Then, $I^2/\sim$ is homeomorphic to the unit sphere $S^2$.

Here, we will consider topology space $X$ as follows
\begin{equation}
  X = \cup \{C_0 + u | u\in \Lambda''\}
\end{equation}
in $\mathcal{M}^{D+1}$ spacetime and define an equivalence relation $~$ on $X$ by
\begin{equation}
  (x^1, x^2) \sim (x^1-a, x^2+h) \,,
\end{equation}
then $X/\sim$ with this quotient topology is homomorphic to helix topology. Here $\Lambda''$ and the unit cylinder cell $C_0$ are given by \cite{Feng:2010qj}\cdash\cite{Feng:2011}
\begin{equation}\label{sub2}
   \Lambda'' = \left\{ ~  n(\mathbbm{e}_1 + \mathbbm{e}_2) ~|~ n \in \mathcal{Z} ~\right\} \,.
\end{equation}
and
\begin{eqnarray}
 \nonumber
   C_0 &=& \bigg\{\sum_{i=0}^{D}x^i \mathbbm{e}_i ~|~ 0\leq x^1 < a,
 -h\leq x^2 < 0 , \\ && -\infty <x^0<\infty, -\frac{L}{2} \leq x^T\leq \frac{L}{2}\bigg\} \,,\label{cell}
\end{eqnarray}
where $T = 3,\cdots, D$. This topology causes the helix boundary condition for a Hermitian massless
scalar field
\begin{equation}\label{helxi boundary condition}
   \phi(t, x^1 + a, x^2, x^T) =  \epsilon \phi(t, x^1 , x^2+h, x^T) \,,
\end{equation}
where $\epsilon$ is a constant that takes $+1$ or $-1$. Here, one can see that if $a=0$ or $h=0$, it returns to the periodicity ($\epsilon =1$) or anti-periodicity ($\epsilon = -1$) boundary condition¡£ In the following, we will focus on the case of $\epsilon =1$. For the case of $\epsilon =-1$, see Ref.\refcite{Feng:2011}.

In calculations on the Casimir effect, extensive use is made of eigenfunctions and eigenvalues of the corresponding
field equation. A Hermitian massless scalar field $\phi(t, x^\alpha, x^T)$ defined in a  (D+1)-dimensional flat
spacetime satisfies the free Klein-Gordon equation:
\begin{equation}\label{eom}
    \left(\partial_t^2 - \partial_i^2\right)\phi(t, x^\alpha, x^T) = 0 \,,
\end{equation}
where $i=1,\cdots, D; \alpha=1,2; T=3,\cdots, D$. Under the boundary condition (\ref{helxi boundary condition}), the
modes of the field are then
\begin{equation}\label{modes}
    \phi_{n}(t, x^\alpha, x^T)= \mathcal{N} e^{-i\omega_nt+ik_x x+ik_z z + ik_Tx^T }\,,
\end{equation}
where $\mathcal{N}$ is a normalization factor and $x^1=x, x^2=z$, and we have
\begin{equation}\label{energy}
    w_n^2 = k_{T}^2 + k_x^2 + \left( -\frac{2\pi n}{h}+\frac{k_x}{h}a \right)^2 = k_{T}^2 + k_z^2 + \left( \frac{2\pi n}{a}+\frac{k_z}{a}h
    \right)^2 \,.
\end{equation}
Here, $k_x$ and $k_z$ satisfy
\begin{equation}\label{kxkz}
    a k_x - hk_z = 2n\pi\,, (n=0,\pm1,\pm2,\cdots) \,.
\end{equation}
In the ground state (vacuum), each of these modes contributes an energy of $w_n/2$. The energy density of the field is
thus given by
\begin{eqnarray}
\nonumber
  &E^{D+1}& = \frac{1}{2 a}
  \int \frac{d^{D-1}k}{(2\pi)^{D-1}} \sum_{n=-\infty}^{\infty} \sqrt{k_T^2 + k_z^2 + \left( \frac{2\pi n}{a}+\frac{k_z}{a}h
    \right)^2  } \,, \\&&\label{tot energy}
\end{eqnarray}
where we have assumed $a\neq 0$ without losing generalities.

\section{Evaluation of the Casimir energy}\label{sec:casimir}

\subsection{Massless scalar field in $2+1$ dimension}
\begin{figure}[h]
\begin{center}
\includegraphics[width=0.4\textwidth]{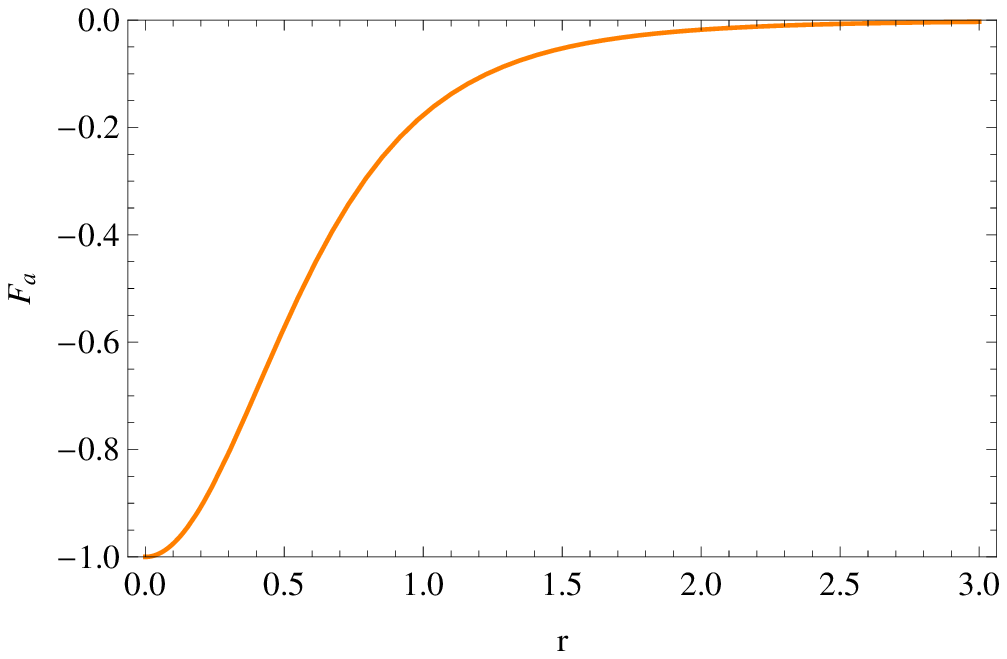}
\qquad
\includegraphics[width=0.4\textwidth]{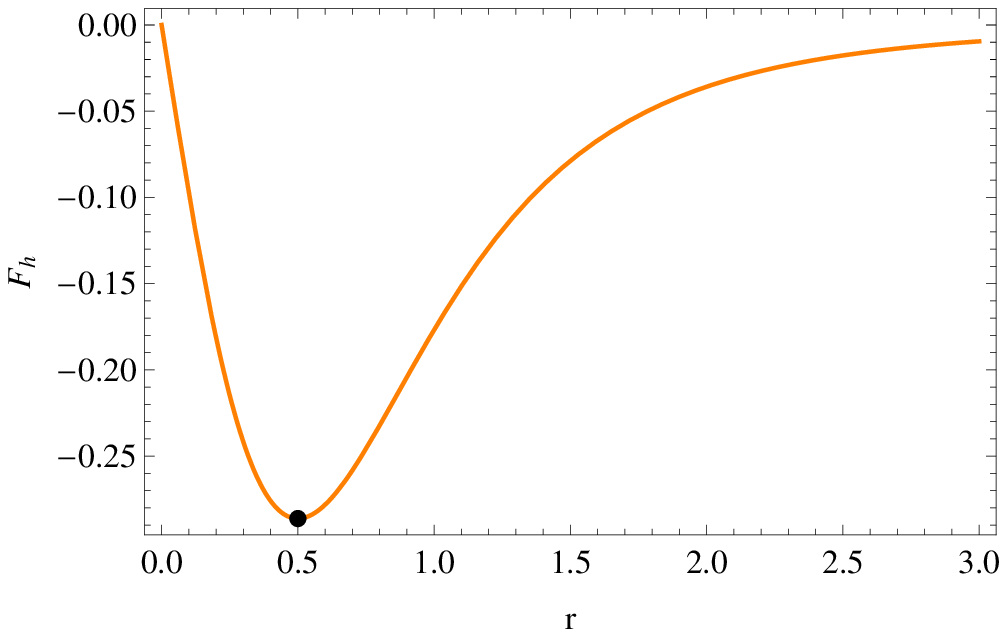}
\caption{\label{fig::force2d}The Casimir force on the $x$ (left) and $z$ (right) direction in the unit $3\zeta(3)/(2\pi
a^4)$ \textit{vs.} the ratio $r$ in $2+1$ dimension. The point corresponds to the maximum magnitude of the force at
$r=0.5$.}
\end{center}
\end{figure}

In the $2+1$ dimensional spacetime, we have the following boundary condition to mimic the helix structure:
\begin{equation}\label{bdc}
   \phi(t, x+a, z)= \phi(t,x,z+h)\,,
\end{equation}
where $h$ is regarded as the pitch of the helix, and we call this condition the helix boundary condition. One can see
from eq.(\ref{bdc}) that it would return to the cylindrical boundary conditions when $h$ vanishes and for $h\neq 0$,
the whole system(the spring) does not have the cylindrical symmetry. Therefore, the vacuum energy  density is given by
\begin{equation}
E(a,h) = \frac{1}{2a}\int_{-\infty}^{\infty} \frac{dk}{2\pi} \sum_{n=-\infty}^{\infty}\sqrt{k^2 + \bigg(\frac{2\pi
n}{a}+\frac{k}{a} h\bigg)^2} \,,
\end{equation}
which is divergent, so we should regularize it to get a finite result. There many regularization method could be used
to deal with the divergence, but in this paper we would like to use the zeta function techniques, which is a very
useful and elegant technique in regularizing the vacuum energy. To use the $\zeta$-function regularization, we define
$\mathcal{E}(s)$ as
\begin{equation}\label{es}
    \mathcal{E}(a,h;s) = \frac{\sqrt{\gamma}}{\pi a}\sum_{n=1}^{\infty}\int_{0}^{\infty} dk\left(k^2 + 1\right)^{-s/2}\left(\frac{2\pi
    n}{a\gamma}\right)^{1-s} \,,
\end{equation}
for $Re(s)>1$ to make a finite result provided by the $k$ integration, and here we have defined
\begin{equation}\label{gamma}
     \gamma \equiv 1+ r^2 \,, \quad  r = \frac{h}{a}\,.
\end{equation}
We will see in the following that the analytic continuation to the complex $s$ plane is well defined at $s=-1$. Thus,
the regularized Casimir energy density is $E_R(a,h)=\mathcal{E}(a,h;-1)$. After integrating $k$ in eq.(\ref{es}), we get
\begin{equation}\label{es2}
   \mathcal{E}(a,h;s) =  \frac{1}{2 a}\sqrt{\frac{\gamma}{\pi}}\left(\frac{2\pi }{a\gamma}\right)^{1-s}
    \frac{ \Gamma\left(\frac{s-1}{2}\right)}{ \Gamma\left(\frac{s}{2}\right)}
    \zeta(s-1) \,,
\end{equation}
where $\zeta(s)$ is the Riemann zeta function. The value of the analytically continued zeta function can be obtained
from the reflection relation
\begin{equation}\label{rel}
    \Gamma\left(\frac{s}{2}\right)\zeta(s) = \pi^{s-\frac{1}{2}} \Gamma\left(\frac{1-s}{2}\right)\zeta(1-s)\,.
\end{equation}
Taking $s=-1$, we get
\begin{equation}
   \lim_{s\rightarrow-1} \Gamma\left(\frac{s-1}{2}\right)\zeta(s-1) =  \frac{\zeta(3)}{2\pi^2}\,,
\end{equation}
then we have
\begin{equation}\label{r e}
    E_R(a,h)  = -\frac{\zeta(3)}{2\pi a^3}\gamma^{-3/2} =-\frac{\zeta(3)}{2\pi a^3}\bigg(1+r^2\bigg)^{-3/2} \,,
\end{equation}
where we have used $\Gamma(-1/2) = -2\sqrt{\pi}$ and if $r=0$, it come back to the cylindrical case  with periodical
boundary, see eq.(\ref{bdc}). The Casimir force on the $x$ direction of the helix  is
\begin{equation}\label{force1 a}
    F_a = -\frac{\partial E_R(a,h)}{\partial a} = - \frac{3\zeta(3)}{2\pi a^4} \bigg(1+r^2\bigg)^{-5/2},
\end{equation}
which is always an attractive force and the magnitude of the force monotonously decreases with the increasing of the
ratio $r$. Once $r$ becomes large enough, the force can be neglected. While, the Casimir force on the $z$ direction is
\begin{equation}\label{force1 h}
    F_h = -\frac{\partial E_R(a,h)}{\partial h} = - \frac{3\zeta(3)}{2\pi a^4} \frac{r}{(1+r^2)^{5/2}},
    \,.
\end{equation}
which has a maximum magnitude at $r =0.5$. When $r<0.5$, the magnitude of the force increases with the increasing of
$r$ until $r=0.5$, and the force is almost linearly depending on $r$ when $r\ll1$. So, it is just like the force on a
spring complying with the Hooke's law, but in this case, the force  originates from the quantum effect, namely, the
Casimir effect. Once $r>0.5$, the magnitude of the force decreases with the increasing of $r$.  To illustrate the
behavior of the Casimir force in this case, we plot them for each direction in Fig.\ref{fig::force2d}.

It should be noticed that in  Fig.\ref{fig::force2d}, the behavior of the forces are different with respect to the
ratio $r$, but this dose not conflict with the relation (\ref{energy}), which shows that labeling the axes is a matter
of convention, namely the final result should have the the symmetry of $a\leftrightarrow h$. The reason is the
following, eq.(\ref{r e}) could be rewritten in terms of $a$ and $h$:
\begin{equation}
    E_R(a,h)  = -\frac{\zeta(3)}{2\pi}\bigg(a^2+h^2\bigg)^{-3/2} \,,
\end{equation}
which respects the symmetry of $a\leftrightarrow h$ in deed. And, one can easily see that eqs. (\ref{force1 a}) and
(\ref{force1 h}) are also under this symmetry, if one rewritten these equations as
\begin{eqnarray}
  F_a &=& - \frac{3\zeta(3)}{2\pi} \frac{a}{(a^2+h^2)^{5/2}}\,, \\
  F_h &=&  - \frac{3\zeta(3)}{2\pi} \frac{h}{(a^2+h^2)^{5/2}} \,,
\end{eqnarray}
which are all consistent with the relation (\ref{energy}).

\subsection{Massless scalar field in $3+1$ dimension}

\begin{figure}[h]
\begin{center}
\includegraphics[width=0.3\textwidth]{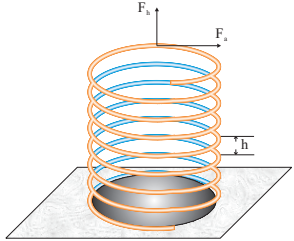}
\caption{\label{fig::spring} Illustration of the \textit{Quantum spring}.}
\end{center}
\end{figure}

\begin{figure}[h]
\begin{center}
\includegraphics[width=0.4\textwidth]{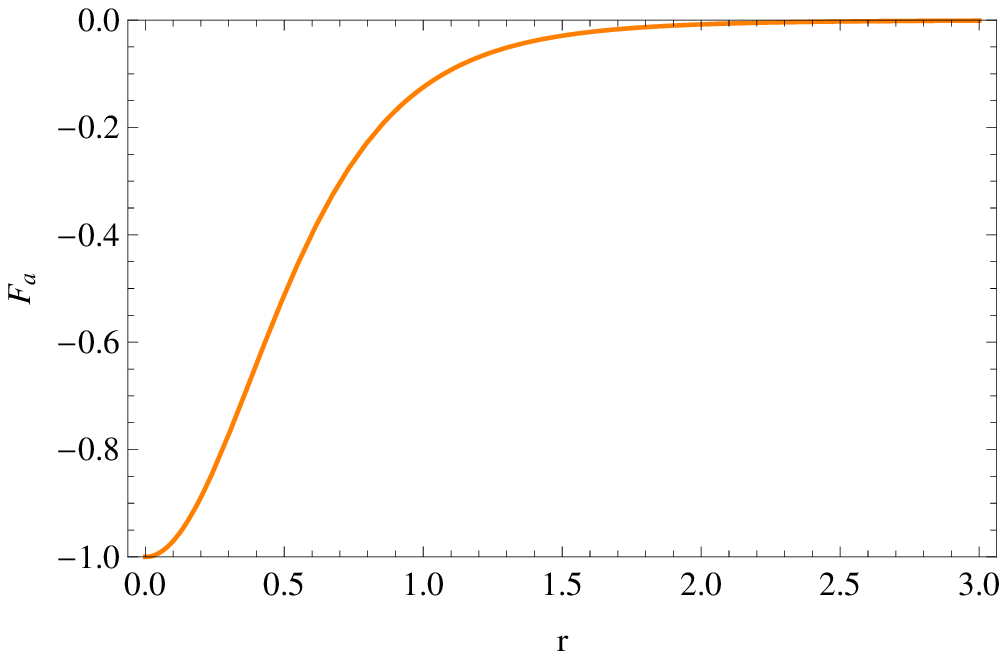}
\qquad
\includegraphics[width=0.4\textwidth]{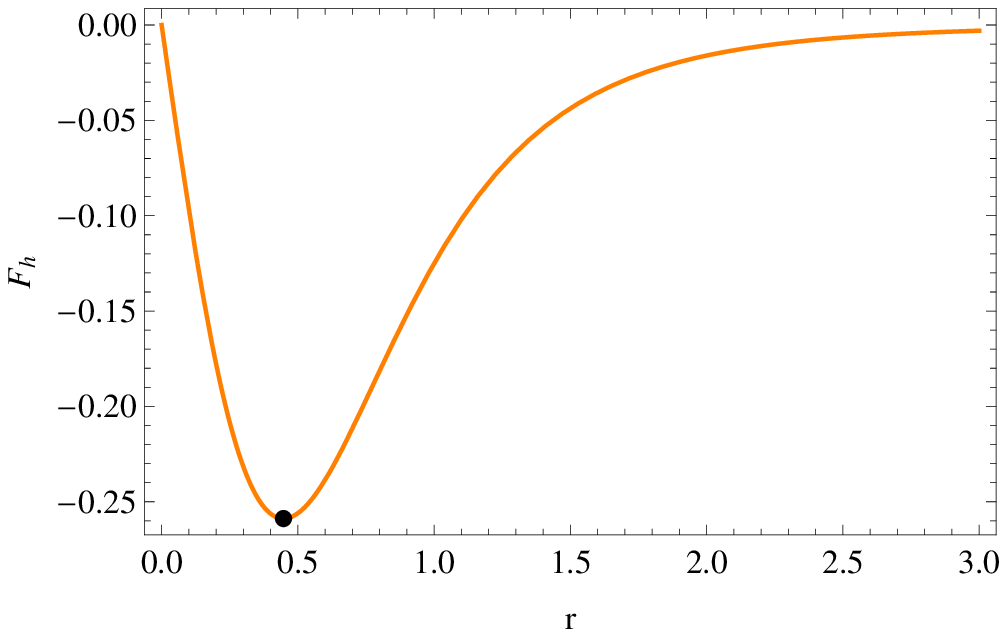}
\caption{\label{fig::force3d}The Casimir force on the $x$ (left) and $z$ (right) direction in the unit $\pi^2/45a^5$
\textit{vs.} the ratio $r$ in $3+1$ dimension. The point corresponds to the maximum magnitude of the force at
$r=1/\sqrt{5}$.}
\end{center}
\end{figure}

As in the $2+1$ dimension case, the vacuum energy  density in $3+1$ dimention is given by
\begin{equation}
E(a,h) = \frac{1}{2a}\int_{-\infty}^{\infty} \frac{dk_ydk_z}{(2\pi)^2} \sum_{n=-\infty}^{\infty}\sqrt{k^2 +
\bigg(\frac{2\pi n}{a}+\frac{k_z}{a} h\bigg)^2} \,,
\end{equation}
where $k^2 = k^2_y + k^2_z$. Again, to use the $\zeta$-function regularization, we define $\mathcal{E}(s)$ as
\begin{equation}\label{es 31}
    \mathcal{E}(a,h;s) = \frac{\gamma}{4\pi a}\sum_{n=1}^{\infty}\int_{0}^{\infty}k dk\left(k^2 + 1\right)^{-s/2}\left(\frac{2\pi
    n}{a\gamma}\right)^{2-s} \,,
\end{equation}
and  for $Re(s)>1$. We will see in the
following that the analytic continuation to the complex $s$ plane is also well defined at $s=-1$ in this case. Thus,
the regularized Casimir energy density is $E_R(a,h)=\mathcal{E}(a,h;-1)$. After integrating $k$  in
eq.(\ref{es 31}), we get
\begin{equation}\label{es2 31}
   \mathcal{E}(a,h;s) = - \frac{\zeta(s-2)}{2\pi(2-s) a}\left(\frac{2\pi}{a \gamma}\right)^{2-s}\,,
\end{equation}
Taking $s=-1$, we get $\zeta(-3)=\frac{1}{120}$ from (\ref{rel}), and then
\begin{equation}\label{r e 31}
    E_R(a,h)  = -\frac{\pi^2}{90 a^4 \gamma^2} =  -\frac{\pi^2}{90 (a^2+h^2)^2}\,.
\end{equation}
Therefore, the Casimir force on the $x$ direction of the helix  is
\begin{equation}\label{force a}
  F_a = - \frac{2\pi^2}{45 a^5} \bigg(1+r^2\bigg)^{-3}
\end{equation}
which is always attractive and its magnitude  monotonously decreases with the increasing of the ratio $r$ and the force will be vanished when $r$ goes to infinity. On
the other hand, the Casimir force on the $z$ direction is
\begin{equation}\label{force h}
    F_h = -\frac{2\pi^2}{45 a^5} \frac{r}{(1+r^2)^3}\,,
\end{equation}
which has a maximum magnitude at $r =1/\sqrt{5}$. And for $r\ll1$, we  have
\begin{equation}
    F_h|_{r\ll1} = -\frac{2\pi^2}{45 a^5}\bigg[r-3r^3 +\mathcal{O}(r^5)\bigg]\,,
\end{equation}
Therefore, for small $r$, the force  linearly depends on $r$, namely,
\begin{equation}
    F_h = - K r \,, \quad K = \frac{2\pi^2}{45a^5}\,,  \quad (r\ll1) \,,
\end{equation}
which is very much like a spring obeying the Hooke's law with spring constant $K$ in classical mechanics, but in this
case, the force comes from a quantum effect, and we would like to call it \textit{quantum spring}, see
Fig.\ref{fig::spring}. To illustrate the behavior of the forces on the helix, we plot them for each direction in Fig.\ref{fig::force3d}.

It should be noticed that this results are also under the symmetry of  $a\leftrightarrow h$ if one rewrite the force in terms of $a$ and $h$ as that in $2+1$ dimensional spacetime.

\section{Conclusion and discussion}

In conclusion, we have investigated the Casimir effect with a helix configuration in two and three dimensions, and it
can be easily generalized to $D+1$ dimensional spacetime and the energy is given by
\begin{equation}\label{fienergy}
E^{D}=-\frac {\pi^{\frac D 2}}{a^{D+1}\gamma^{\frac {D+1}{2}}}\Gamma \left (-\frac D 2 \right )\zeta(-D) \,.
\end{equation}
It should be noticed that when $D$ is odd, one can rewriter the above equation in terms of the Bernoulli numbers, while $D$ is even, one should use the reflect relation (\ref{rel}) to get the final result, for more details see Ref.~\refcite{Zhai:2010mr}. Then, the corresponding Casimir forces can be straightforwardly obtained and one can see that the force parallel to the axis of the helix has a particular behaviors that the Casimir force in the usual case do no possesses. It behaves very much like the force on a spring that obeys the Hooke's law in mechanics.
Furthermore, the The magnitude of this force has a maximum values at $r = 0.5$ (2D) or $r = 1/\sqrt{5}$ (3D). For high dimensions, the critical ratio is at $r = 1/\sqrt{D+2}$.

Therefore, we would like to call this helix configuration as a \textit{quantum spring}, see Fig.\ref{fig::spring}. On the other
hand, the force perpendicular to the axis decreases monotonously with the increasing of the ratio $r$. Both forces are
attractive and their behaviors are the same in two and three or even higher dimensions.

In this paper, we have considered the massless scalar field, and one can easily generalize it to a massive scalar
field or a massive spinor \cite{Feng:2011}. As is known that the Casimir effect disappears as the mass of the field goes to infinity since there are no more
quantum fluctuation in this limit, but of course, how the Casimir force varies as the mass changes is worth studying
\cite{Bordag:2001qi}. Since this \textit{quantum spring} effect may be
detected in the laboratory and be applied to the microelectromechanical system, we suggest to do the experiment to
verify our results. It should be noticed that, in the experiment or the real application, the spring  like
Fig.\ref{fig::spring} should be soft, which means the force coming from the classical mechanics must be small enough,
and the quantum effect dominates the behavior of the spring.

\section*{Acknowledgments}
This work is supported by National Science Foundation of China grant No.~11047138, National Education Foundation of China grant  No.~2009312711004 , Shanghai Natural Science Foundation, China grant No.~10ZR1422000 and  Shanghai Special Education Foundation, No.~ssd10004.



\begin{thebibliography}{0}    

\bibitem{Casimir:1948dh}
  H.~B.~G.~Casimir,
 {\it Indag.\ Math.}  {\bf 10}, 261 (1948)
  [{\it Kon.\ Ned.\ Akad.\ Wetensch.\ Proc.\ } {\bf 51}, 793 (1948\ FRPHA,65,342-344.1987\ KNAWA,100N3-4,61-63.1997)].

\bibitem{Plunien:1986ca}
  M.~Bordag, G.~L.~Klimchitskaya, U.~Mohideen and V.~M.~ Mostepanenko, \textit{Advances in the Casimir Effect}, Oxford
  University Press, 2009.


\bibitem{Utiyama:1962sn}
  R.~Utiyama and B.~S.~DeWitt,
  {\it J.\ Math.\ Phys.\  } {\bf 3}, 608 (1962).

\bibitem{DeWitt:1975ys}
  B.~S.~DeWitt,
  {\it Phys.\ Rept.\  } {\bf 19}, 295 (1975).

\bibitem{Decca:2007yb}
  R.~S.~Decca, D.~Lopez, E.~Fischbach, G.~L.~Klimchitskaya, D.~E.~Krause and V.~M.~Mostepanenko,
  {\it Phys.\ Rev.\  D }{\bf 75}, 077101 (2007)

\bibitem{MEMS}
  F.~M.~Serry, D.~Walliser, and G.~J.~Maclay, {\it J.Microelectromech.Syst. } {\bf4}, 193 (1995), \\
  H.~B.~Chan, V.~A.~Aksyuk, R.~N.~Kleiman, D.~J.~Bishop, and F.~Capasso, {\it Science } {\bf291}, 1941 (2001).

\bibitem{Emig:2007cf}
  T.~Emig, N.~Graham, R.~L.~Jaffe and M.~Kardar,
  {\it Phys.\ Rev.\ Lett.\  }{\bf 99}, 170403 (2007)

\bibitem{Lukosz}
  W.~Lukosz, {\it Physica }{\bf 56}, 109(1971).

\bibitem{Li}
  X.~Z.~Li, H.~B.~Cheng, J.~M.~Li and X.~H.~Zhai,
  {\it Phys.\ Rev.\  D }{\bf 56}, 2155 (1997).

\bibitem{Li2}
  X.~Z.~Li and X.~H.~Zhai,
  {\it J.\ Phys.\ A }{\bf34}:11053-11057, 2001.


\bibitem{Elizalde}
  E.~Elizalde, S.~D.~Odintsov, A.~Romeo, A.~A.~Bytsenko and S.~Zerbini, \textit{Zeta Regularization Techniques with
  Applications}, World Scientific, Singapore, 1993.

\bibitem{Helliwell:1986hs}
  T.~M.~Helliwell and D.~A.~Konkowski,
  {\it Phys.\ Rev.\  D }{\bf 34}, 1918 (1986).

\bibitem{Li:1990bz}
  X.~Z.~Li, X.~Shi and J.~Z.~Zhang,
  {\it Phys.\ Rev.\  D }{\bf 44}, 560 (1991).

\bibitem{Li:1990bz2}
  I.~H.~Brevik, H.~B.~Nielsen and S.~D.~Odintsov,
  {\it Phys.\ Rev.\  D }{\bf 53}, 3224 (1996).

\bibitem{BezerradeMello:1999ge}
  E.~R.~Bezerra de Mello, V.~B.~Bezerra and N.~R.~Khusnutdinov,
  {\it Phys.\ Rev.\  D }{\bf 60}, 063506 (1999).

\bibitem{Shi:1991qc}
  X.~Shi and X.~ .~Li,
  {\it Class.\ Quant.\ Grav.\  }{\bf 8}, 75 (1991).

\bibitem{Zhai}
  X.~H.~Zhai and X.~Z.~Li,
  {\it Phys.\ Rev.\  D }{\bf 76}, 047704 (2007).

\bibitem{Zhai2}
  X.~H.~Zhai, Y.~Y.~Zhang and X.~Z.~Li,
  {\it Mod.\ Phys.\ Lett.\  A }{\bf 24}, 393 (2009).

\bibitem{Zhai3}
  R.~M.~Cavalcanti,
  {\it Phys.\ Rev.\  D }{\bf 69}, 065015 (2004).

\bibitem{Zhai4}
  M.~P.~Hertzberg, R.~L.~Jaffe, M.~Kardar and A.~Scardicchio,
  {\it Phys.\ Rev.\ Lett.\  }{\bf 95}, 250402 (2005).

\bibitem{Feng:2010qj}
  C.~J.~Feng and X.~Z.~Li,
  {\it Phys.\ Lett.\  B }{\bf 691}, 167 (2010).

\bibitem{Zhai:2010mr}
  X.~H.~Zhai, X.~Z.~Li and C.~J.~Feng,
  {\it Mod.\ Phys.\ Lett.\  A }{\bf 26}, 669 (2011).
  [arXiv:1008.3020 [hep-th]].

\bibitem{Feng:2011}
  X.~H.~Zhai, X.~Z.~Li and C.~J.~Feng,
  {\it Eur. Phys. J. C } {\bf 71 }, 1654 (2011).

\bibitem{Mukres}
  J.~R.~Munkres, {\it Elements of Algebraic Topology}, Addison-Wesley Publishing Company, Amsterdam, 1984.

\bibitem{integ}
I.S. Gradshteyn and I.M. Ryzhik ; Alan Jeffrey, Daniel Zwillinger, editors. \textit{Table of Integrals, Series, and
Products}, seventh edition. Academic Press, 2007. ISBN 978-0-12-373637-4 .

\bibitem{Bordag:2001qi}
  M.~Bordag, U.~Mohideen and V.~M.~Mostepanenko,
  {\it Phys.\ Rept.\  }{\bf 353}, 1 (2001).

\end{thebibliography}
\end{document}